\documentstyle[12pt,epsfig]{article}
\def\be{\begin{equation}}
\def\ee{\end{equation}}
\def\bea{\begin{eqnarray}}
\def\eea{\end{eqnarray}}
\def\half{\frac{1}{2}}
\def\fn{\footnote}
\def\bh{black hole \ }
\def\bhs{black hole solutions \ }
\def\cL{{\cal L}}
\def\cH{{\cal H}}

\def\mn{{\mu\nu}}
\def\m{\mu}
\def\n{\nu}
\def\a{\alpha}
\def\b{\beta}

\def\Br{Bronnikov {}}
\def\AG{Ay\'on-Beato -- Garc\'\i a {}}

\def\E{\bf E}
\def\B{\bf B}
\def\H{\bf H}
\def\D{\bf D}
\def\d{\partial}

\begin{document}

\title{New Type of Regular Black Holes and Particlelike Solutions from
Nonlinear Electrodynamics}

\author{Alexander Burinskii\\
Gravity Research Group, NSI Russian
Academy of Sciences,\\
B. Tulskaya 52, 113191 Moscow, Russia
\thanks{E-mail:bur@ibrae.ac.ru}\\
\&\\ Sergi R. Hildebrandt
\thanks{Temporary address: Avda. Mar\'\i tima, 49, P041E. Candelaria, 38530.
S/C. de Tenerife. Spain. E-mail: hildebrandt@ieec.fcr.es} \\
Institut d'Estudis Espacials de Catalunya (IEEC/CSIC)\\
Edifici Nexus, Gran Capit\`a 2-4, 08034 Barcelona, Spain\\
\\}

\date{E-print: hep-th/0202066}

\maketitle

\begin{abstract}
\noindent
We show that the Bronnikov theorem
on the nonexistence of regular electrically charged black holes
can be circumvented. In the frame of nonlinear electrodynamics, we
present the exact regular black hole solutions of a hybrid type.
They are electrically charged, but contain a `dual' core confining
a polarization of magnetic charges.

     The considered example is based on a modification of the
Ay\'on-Beato \& Garc\'\i a solution. It represents a very
specific realization of the idea on confinement based on dual
electrodynamics and dual superconductivity.


\end{abstract}

\section{Introduction}
Last years considerable interest in regular  black hole
solutions has been renewed  \cite{behm,Dym,Bag}\fn{See also list of
references in
\cite{behm}.}, and in particular, to those based on nonlinear
electrodynamics
(NED) \cite{ag1,ag2,ag3,Br,nov}.  Analyzing  the Ay\'on-Beato \& Garc\'\i a
black hole solutions from nonlinear electrodynamics \cite{ag1,ag2,ag3},
Bronnikov presented three important results \cite{Br}:

(i) - a version of no-go theorem claiming the
nonexistence of the regular electrically charged black holes and
particlelike solitons with regular center
\fn{As well as wormholes or horns.};

(ii) - the existence of cusps and branches in the Lagrangian used for the
\AG
solution \cite{ag2}; and

(iii) - a class of exact regular magnetically charged black holes and
solitonic solutions.

He showed that  the branch, realized in the \AG solution near the core, does
not tend in the weak field limit to the Maxwell theory and presented two
theorems leading to the
conclusion that regular \bh and solitonic solutions can be only
magnetically charged.

The unusual properties of the \AG  solution obtained by \Br
have captured our attention in connection with recent suggestions on a phase
transition which has to occur for regular \bh  and particlelike solutions in
the core region \cite{behm,Dym,Bag}. In this work we show that the
conditions of the Bronnikov theorem can be circumvented by dealing with
the models having a phase transition near the core of solution, when
electric
field does not extend to the center of the solution.

After some modification of the \AG  solutions, we obtain a new type
of regular, electrically charged black hole solutions. The main
peculiarity
of these solutions is a `magnetic' core described by dual electrodynamics.
This class of solutions deserves interest not only as an explicit example
of the regular electrically charged \bh solution, but also as an example
of particlelike solution with a very
specific realization of the ideas on quark confinement based on dual
electrodynamics \cite{Dir} and dual superconductivity
\cite{dual,SW,Ch,MS,KIK}.

\section{NED and F-P Dual representations }

    The action for NED in general relativity is
\be
S = \frac{1}{16\pi} \int d^4 x\sqrt{-g} [R - \cL(F)],   \label{S}
\ee
where $R$ is the scalar curvature, $F=F_\mn F^\mn$,
$F_\mn = \d_\m A_\n - \d_\n A_\m,$ and $\cL (F)$ is a nonlinear function.

The equations, following from Eq. (\ref{S}) are
\be
\nabla_\m (L_F F^\mn)=0,\qquad
\nabla_\m \star F^\mn =0.    \label{Feq}
\ee
where $L_F = d\cL/dF$,
and the equations $\nabla_\m \star F^\mn =0 $
are the Bianchi identities.
\fn{We use signature $-+++$.
The Hodge star operator is given by
$\star F^\mn = \half\eta^{\mn \a\b} F_{\a\b}$,
$\eta_{\mn \a\b}$ is completely skew-symmetric,
$\eta_{0123}=\sqrt{-g}$.}

By introducing the electric intensity $\E$ and magnetic induction $\B$,
\be
{\E} = \{ F_{i0}\}, \quad
{\B} = \{\star F_{i0}\}= \half \eta_{i0}^{\ \ \a\b} F_{\a\b},
\label{EB}
\ee
one can obtain from Bianchi identities $\nabla _\m \star F^\mn =0$
the pair of Maxwell equations
$\nabla {\B}=0; \qquad \nabla \times {\E}=-\partial{\B}/\partial t .$
The dynamic equations $\nabla_\m (L_F F^\mn)=0 $ can be expressed via  the
tensor
\be
P^\mn = \partial \cL /\partial F_\mn = L_F F^\mn,
\label{P}
\ee
in the form
\be
\nabla _\m P^\mn =0;
\label{eqP}
\ee
that, in the terms of the electric induction $\D$
and magnetic intensity $\H$,
\be
 {\D} =\{P_{i0}\} = L_F \{F_{i0}\}, \quad
{\H} =\{\star P^{0i}\} = L_F \{\star F^{0i}\}, \label{DH}
\ee
yields the second pair of the Maxwell equations
$\nabla \D=0; \qquad \nabla \times \H=\partial\D/\partial t .$
These equations show that in NED $L_F$ plays the role of
electric susceptibility $\epsilon = L_F$ and the magnetic permeability
is given by $\mu = 1/L_F$.

Relation (\ref{P}) allows one to rewrite the field equations (\ref{Feq})
in an equivalent dual P form
\be
\nabla_\m (P^\mn )=0, \qquad \nabla_\m \frac 1 {L_F} \star P^\mn =0.
\label{Peq}
\ee
It should be mentioned that it is a dual description of {\it the same
physical system}.
This description can be obtained from the electromagnetic
Lagrangian  $\cH(P)$ ( expressed via  $P= P_\mn P^\mn= L_F^2 F$ )
determined by Legendre transformation \cite{ag2,Br,SGP}
\be
\cH(P) =2F L_F -\cL.
\label{Leg}
\ee
The following relations can be easily obtained
\be
\cL=2P H_P -\cH; \qquad H_P= d \cH /d P = 1/ L_F.
\label{FPd}
\ee
{\it The Hodge star duality} operation allows one to give a
description of the field intensities via dual components
$\tilde F_\mn =\star F_\mn = \half\eta_\mn^{\ \ \a\b} F_{\a\b}$.
Since $\star\star=-1$, by using the Hodge dual description
 the following relations hold:
\be
F  = - \tilde F; \qquad P=-\tilde P;
\label{FPtilde}
\ee
and also
\be
 L_F= - L_{\tilde F};\qquad H_P= - H_{\tilde P};\qquad
\tilde L(\tilde F)=L(-F);\qquad\tilde \cH (\tilde P)=\cH(-P);
\label{Ltilde}
\ee
and the equations (\ref{Feq}) take in this description the form
\be
\nabla_\m (L_F \star \tilde F^\mn)=0, \qquad \nabla_\m \tilde F^\mn =0.
\label{Eqtilde}
\ee
Note that the Hodge dual description changes sign of $L_F$ that can lead
to nonphysical values of electric susceptibility $\epsilon  <0$
\fn{See, for example, \cite{LL}}.

Vice versa, if the Lagrangian leads to
$L_F<0$ it can be considered as a sign to pass to another, physically
valid (Hodge) dual description.

\section{Nonexistence of Electrically Charged Regular Black Holes?}

The Bronnikov no-go theorem is based on the assumption that the NED
Lagrangian has to lead to Maxwell theory at small $F$: $\cL(F) \to F$
as $F\to 0$. This is equal to $L_F \to 1$ at small $F$. He shows that real
behavior of electrically charged solution near the center leads to
finite values of $|FL_F|$ and to divergence of $|F|L_F^2$. Therefore at
the center the ratio $\frac {|F|L_F^2}{FL_F}\equiv L_F $ tends to infinity
that is non-Maxwellian behavior.

 This derivation is correct. However, there are two points which throw
doubts on the initial postulates. The first one is connected
with the assumption that  the electromagnetic field is
extended up to the center of the solution. At least in flat space there is a
remarkable example of the electrically charged solution, present only in
the region $r\ge a$. It is the classical Dirac electron model, as a charged
sphere. \fn {The Bronnikov theorem 2 tries to exclude this case showing
nonexistence of a horn in space-time. However, there can be other
(nonmetrical) mechanisms restricting the penetration of the electromagnetic
field into region $r<a$, in particular, superconducting properties of
core or a special behavior in NED as it will be shown further.}
The second argument is connected with the demands of Maxwellian behavior.
Indeed, the condition $ L_F \to 1 $ as $ F \to 0 $
(to recover Maxwell theory) is reasonable at spatial infinity.
However, at the core, although
the gravitational field is regular everywhere and becomes {\em weaker} as
one approaches the center (the metric tends to flat space-time), the spatial
volume is much smaller (on many orders) than in the asymptotic limit
where gravity is again weak and where $ L_F \to 1 $ makes sense.
Thus both types of boundaries are essentially different.
 Another argument can be a strong vacuum
polarization and the presence of other ( maybe virtual quantum ) fields
which
have a straight relation to the formation the nonvacuum values of
$L_F =\epsilon$.
Therefore $\cL(F)$ can have a nonlocal behavior
depending not only on the value of $F$ but also on full field configuration
and
proximity to the core.

  If one restricts oneself to Lagrangian densities which are {\em local
functions} of $ F $, then, Bronnikov is completely right. But
assigning essentially the same physics to the core region as to the
asymptotic region (in electromagnetic terms) is not convincing enough.
Therefore his restrictions for regular solutions are too especial to be
accepted in general. This point of view is suggested in fact by the
given Bronnikov analysis showing the existence of cusps
and branches in the Lagrangian of the \AG  solution.
In fact, his analysis shows that the \AG Lagrangian has different
behavior at the core and far outside the body, although $ F \to 0 $ in both
cases.

\section{The \AG Solution}

The electrically charged regular \bh solution has been given by
\AG in \cite{ag2}. Most investigations in nonlinear electrodynamics
are connected with the Born-Infeld type of nonlinearity (see, e.g.,
\cite{gibras,gibras2,Her,Ts} and references therein). However, as it was
mentioned for regular \bhs in \cite{ag2}, the ``...Born-Infeld Lagrangian
is useless in this topic, since it gives rise to a singular black
hole solution, at least in the case of spherical symmetry \cite{HI}...'',
and \AG start from the nonlinear function
\be
\cH (P) =P/\cosh^2\{s(-2q^2P)^{1/4}\}, \label{AGH} \ee
where $s=|q | /2m $
($ q$ is the electric charge and $m$ is the mass of the body).
\begin{figure}[ht]
\centerline{\epsfig{figure=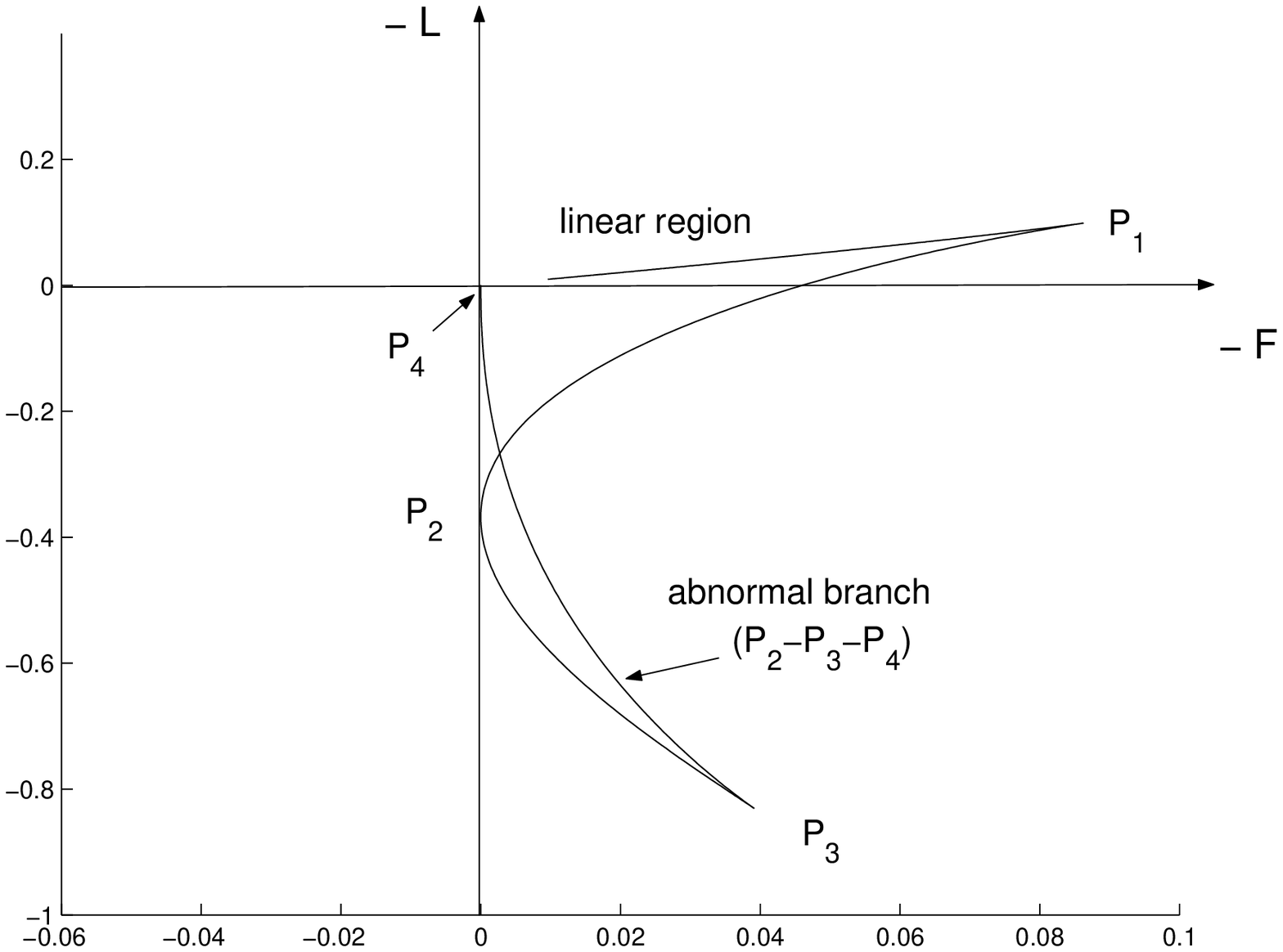,height=7cm,width=9cm}}
\caption{\small Dependence $\cL (F)$ for \AG solution.}
\end{figure}

Note, that particular form of nonlinearity $\cH (P)$ is not essential,
and existence of regular \bhs is determined by the asymptotic
properties of this function.
What is important is the Maxwellian behavior on large distances,
$\cH (P) \equiv P $ at small $P$,
and a rapid fall-off of this function by $ P\to\infty $.
These demands lead to the existence of an extremum of the
function $\cH $, and moreover to the existence of at least one sector where
$H_P$ is {\it negative}.
For instance, we use for illustration the simplest function of this sort
\be
\cH(P)= P\exp\{-|P|\}.
\label{HP}
\ee
The corresponding Lagrangian in L description is
$\cL(F) = 2H_P P - \cH(P)$, where $P$ has to be expressed via
$F=P/L_F^2$, ( see Fig. 1 and upper graph in Fig. 2.) .

It yields the system of electromagnetic equations
\be
\nabla_\m (P^\mn )=0;
\label{Peq1}
\ee
and Einstein equations $G_\m^\n = T_\m^\n$, where the energy-momentum
tensor is given by
\be
T_\m^\n = \half (4L_F F_{\m\a} F^{\n\a} -\delta _\m^\n \cL).
\label{T}
\ee

In the case of spherical symmetry the electrically charged solution of
Eq. (\ref{Peq1}) is
\be
P_{0r} = - P_{r0}= q/r^2
\label{P0r}
\ee
or
\be
F_{0r} = - F_{r0}=L_F^{-1} q/r^2;
\label{F0r}
\ee
the other components being equal to zero.
Thus $P=-2q/r^4.$
\begin{figure}[ht]
\centerline{\epsfig{figure=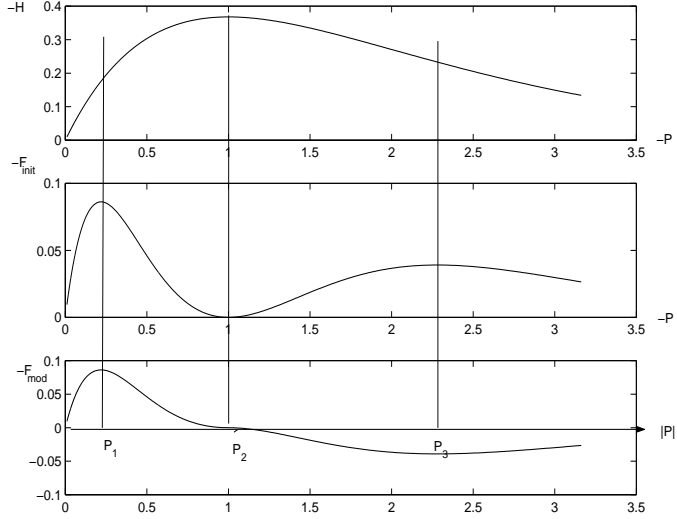,height=7cm,width=9cm}}
\caption{\small Graphs of $\cH (P)$ and $F(P)$ for \AG solution and modified
solution $F_{mod}(P)$.}
\end{figure}
The Einstein equations can easily be solved leading to the mass term
$M(r)= m [ 1-\tanh(q^2/2mr)]$ and to the metric
\be
ds^2 = -[1-2M(r)/r ]dt^2 + [1-2M(r)/r ]^{-1} dr^2 + r^2 d\Omega^2,
\label{metric}
\ee
which is regular by $r$ going to zero.

As mentioned by Bronnikov \cite{Br}, the Lagrangian given by the smooth and
regular function $\cH(P)$ displays  cusps and branches in F description.
Appearance of cusps at the points $P_1$ and $P_3$  is connected with
multivaluedness of the inverse map  $P(F)$ ( see fig.2) and is a general
feature of the considered class of nonlinear functions.
Very specific behavior of function $L(F)$ by large $P$, on the section
$P_2 - P_3 - P_4 $ is determined by the following phenomenon:
The magnetic permeability $\mu =H_P$ is going to zero as
$P\to P_2$, the point of $\cH$ extremum. Consequently, electric
susceptibility $\epsilon =1/H_P$ is divergent at this point. The sphere,
determined by the equation $P=P_2$ looks like an ideal conducting surface,
and electromagnetic field $F_\mn $ ( and $F$) will tend to zero approaching
this surface.  The solution, analytically extended inside this sphere
departs drastically from a Maxwellian behavior.
Nevertheless it still has reasonable properties in many respects. For
instance, the electromagnetic field $F_\mn$ and the invariant $F$ go both to
zero as the very center of the solution is approached ( although $L_F \to
\infty$ at $P_2$ and $P_4$) and the metric is regular everywhere, leading
to regular \bhs.

In spite of these properties, the physics inside the core looks rather
unusual since the value of electric susceptibility $\epsilon =1/H_P$ is
negative there and does not correspond to any known physical mediums
\cite{LL}.  Combining NED and general relativity, one can show
from the system of equations (\ref{Peq1}), (\ref{T}) that
the corresponding contribution of effective charges to the metric
will be of opposite sign in the regions with negative $L_F $,
as if the charge would be imaginary valued there.

\section{Modification of the \AG Solution}

The fact that regularity of \bhs can be reached without troubles in
the magnetically charged solutions \cite{Br} suggests the idea to
combine the electrically and magnetically charged solutions
by using a phase transition near the core to magnetic phase,
regularizing the \bh singularity.  In this case, the existence of
different branches in the \AG solution represents a perfect possibility for
realization of this idea by means of matching external and internal
solutions with necessary properties. As we shall see, the transfer
to dual electromagnetic phase near the core of the \AG solution allows
also one to avoid abnormal branch $(P_2-P_3-P_4)$ with negative
$L_F$.

Looking on the graph $L(F)$ (Fig. 1) one sees that it can only be achieved
by
replacing this branch by the branch obtained by a mirror
map with respect to the $L$ axis, as it is shown in Fig. 3. This replacement
takes place in the region of strong field $|P|>|P_2|$, i.e.,  inside of  the
sphere $r<r_0 = |\frac{2q}{P_2}|^{1/4}$ surrounding the singular point of
$P(r)$.

\begin{figure}[ht]
\centerline{\epsfig{figure=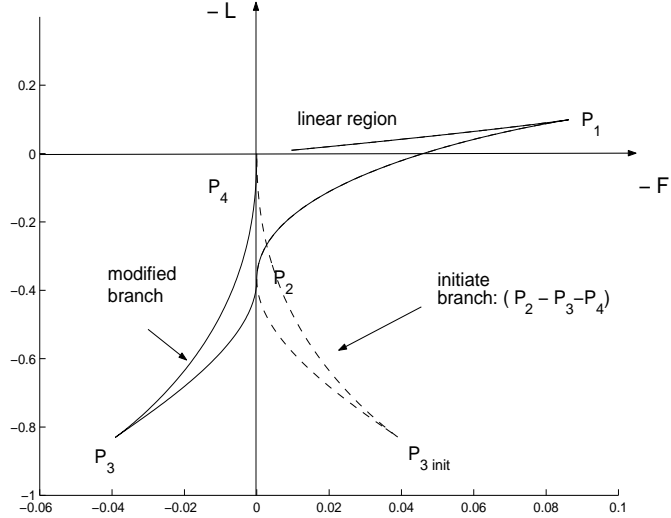,height=7cm,width=9cm}}
\caption{\small Initiate and modified dependencies $L(F)$.}
\end{figure}

As we shall see this replacement corresponds to ``dual"
electrodynamics with a polarization of magnetic charges inside the sphere
$r=r_0$, while the solution for $r>r_0$ survives to be the electrically
charged \AG one. Matching of the modified internal solution and the external
\AG one is smooth with respect to the metric and physical field intensities.

     The properties of modified Lagrangian $\tilde L(\tilde F)$
and the modified relations will be the following.

\begin{enumerate}
 \item  The
region of definition of Lagrangian $\tilde L$  is extended to positive
values
of the invariant $\tilde F = B^2 - E^2$, and modified electric
susceptibility,
$\tilde\epsilon=\tilde L_{\tilde F}= -L_F$, is positive.

\item  Matching of the new branch with the initial one is smooth and occurs
at the point $P_2$ where $F=\tilde F=0$ and
$ L(F)|_{P_2}=\tilde L(\tilde F)|_{P_2}$.

\item
The modified dual intensities are related as
\be
\tilde P ^\mn = \tilde L _{\tilde F} \tilde F ^\mn ; \quad
\tilde P = (\tilde L_{\tilde F} )^2 \tilde F.
\ee
At the point $P_2$ electric susceptibility is divergent
$ L_F|_{P_2}=\tilde L _{\tilde F}|_{P_2} =+\infty $, while the
magnetic permeability $\mu=1/L_F$ is bounded and $C^1$.

\item
As far as in the
modified region $\tilde F>0$ we get also $\tilde P >0$.
On the other hand
$P=H^2-D^2$ is negative in the region $r>r_0$ (as in the case of an
electrically charged solution).  Therefore the modified function
$P(r)$ will be discontinuous.  The change of sign of the invariant $P$ at
the
point $P_2$ is evidence of a transfer to magnetic phase by $r<r_0$.

\end{enumerate}

Due to the Hodge star
operation identity $\star\star = -1$, the equality $\tilde P = -P$ can be
  resolved as  $\tilde P^\mn = \pm \star P^\mn$.  Whence, the relation
(\ref{EB}) shows that the modified \AG solution (\ref{F0r}) by $r<r_0$
describes a field of magnetic induction in radial direction $\bf n $

\be
{\tilde{\B}} = \mp q {\bf n} \tilde L_{\tilde F} ^{-1}/r^2.
\label{Bmod}
\ee
Magnetic induction $\B$ characterizes the true intensity of the field
strength in media \cite{LL}. In modified solution it tends to zero at
$r=0$ and at $r=r_0$.
Contrary to the standard Maxwell equation $\nabla \B =0$ describing
the absence of magnetic charges, we obtain here the magnetic charge
distribution
\be
\nabla \tilde {\B} = J_{mag} = \mp(q/r^2) {\bf n}\nabla \tilde L_{\tilde F}
=\mp (q/r^2) \tilde L_{\tilde F\tilde F}\frac {d\tilde F}{dr} , \label{Jmag}
\ee
displaying realization of dual electrodynamics in the core region and
singular concentration of magnetic charges  at $r=0$ and $r=r_0$.

\begin{figure}[ht]
\centerline{\epsfig{figure=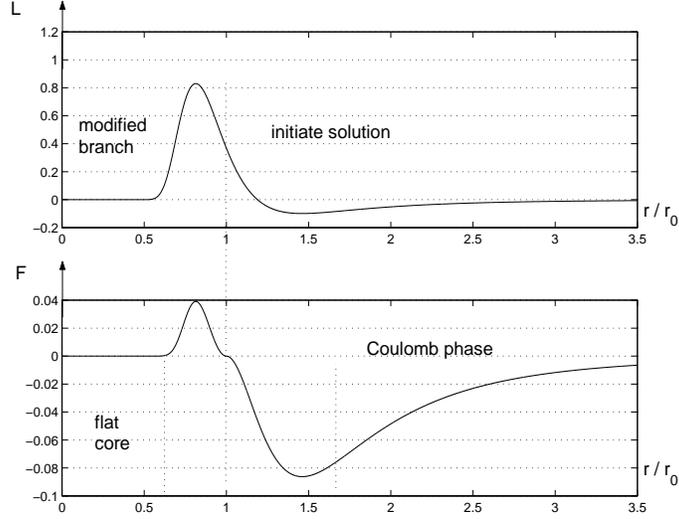,height=7cm,width=9cm}}
\caption{\small Radial dependencies  of $L$ and $F$ for modified solution.}
\end{figure}
The initial and modified dependencies $\cH (P) $, $F(P)$, and $L(F)$ are
given in  figs. 1 and 3.

\subsection{Stress-energy tensor and metric.}

We use the Kerr-Schild form of metric \cite{behm,DKS}.
\fn{See also \cite{EH} where it was shown that the static solutions of the
Einstein equations with spherical symmetry and $T^0_0 = T^1_1$ belong to
the Kerr-Schild class.
\par
In this section we use slightly different units, adapting to the works
\cite{DKS, Bag, behm} and setting $G=1$. The reason for this is the
possibility of using the Kerr-Schild formalism for further extension
to rotating solutions.}
In the case of spherical symmetry
\be
g_\mn=\eta_\mn + [f(r)/r^2] k_\m k_\n , \label{gKS}
\ee
where $\eta_\mn$ is the auxiliary Minkowski metric and
$k_\m = \{1,{\bf n}\}$.
Co- and contravariant forms of the metric in the Kerr angular coordinates
$\{t, r,\theta, \phi \}$ are given in the Appendix.
We have
\be F^{01} =-F_{01} ;\quad
\tilde F_{23}= r^2 \sin \theta F^{01};\quad
\tilde F^{23}= \tilde F_{23}/(r^4 \sin ^2 \theta);
\ee
which, taking into account Eq. (\ref{F0r}), yield
\be \tilde F
= 2 \tilde F^{23} \tilde F_{23}= 2 (F_{01})^2 = 2L_F^{-2}q^2/r^4.  \ee
Similarly, \be \tilde P = 2 \tilde P^{23} \tilde P_{23}= 2 (P_{01})^2 =
2q^2/r^4.  \ee
The modified energy-momentum tensor inside the core acquires
the form
\be T_\m^{(in)\n} = - \half (\delta_\m^\n \tilde L -4 \tilde P_{\m\alpha}
\tilde P^{\n \alpha}/{\tilde L}_{\tilde F}) =
-\half {\it diag} \{\tilde L,\tilde L,\tilde L -2\tilde P /
{\tilde L_{\tilde F}},
\tilde L-2\tilde P /{\tilde L}_{\tilde F} \}.  \label{Tin}
\ee
Outside the core, by $r>r_0$ it is
\be
T_\m^{(ext)\n} = - \half (\delta_\m^\n L -4 P_{\m\alpha}
P^{\n \alpha}/L_F) = -\half{\it diag} \{ L - 2P / L_F, L- 2P /L_F,L,L \},
\label{Tel}
\ee
where $P=-\tilde P= -2q^2/r^4$.
As far as contributions of terms
$2P/L_F$ and $2\tilde P /{\tilde L}_{\tilde F}$, together with their first
derivative, tend to zero approaching $r_0$, one sees that these expressions
match smoothly at $r_0$.

 \begin{figure}[ht]
\centerline{\epsfig{figure=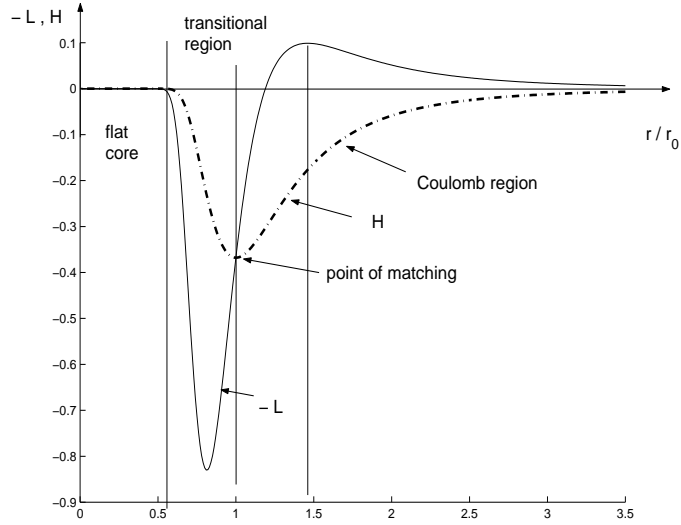,height=7cm,width=9cm}}
\caption{\small Radial dependencies of $-L$ and $\cH$.}
\end{figure}

Comparing corresponding expressions for the generalized
Kerr-Schild class of metrics \cite{behm}
\be
T_\m^\n = (8\pi)^{-1}{\it diag}\{-2G,-2G,D+2G,D+2G\}, \label{TKS}
\ee
where (see the Appendix)
\be
D= -f''/r^2, \quad G= (f/r)'/r^2 \label{DG}
\ee
and $'\equiv d /d r$   gives the relations:

For external field
\be
- \cH _{(ext)} (r)= \frac {(f/r)'}{2\pi r^2} =2 \rho (r), \label{Eext}
\ee
where $\rho(r)$ is the energy density in this region.

For the external part of the function $f(r)$
\be
f(r)=r M_{(ext)}(r),\quad M_{(ext)}(y)
=-2\pi \int_{r_0}^{y}{\cH}_{(ext)}(r)r^2
dr + C,
\label{fext}
\ee
where $C$ is the integration constant.

For function $f(r)$ inside the core,
$r<r_0$, we obtain \be \tilde L (r)= \frac {(f/r)'}{2\pi r^2} =2 \rho (r),
\label{Eint}
\ee
that gives rise for the internal region to
\be
f(r)=r M_{(int)}(r),\quad M_{(int)}(y)=2\pi \int_0^{y}{\tilde L}(r) r^2 dr.
\label{fint} \ee
Each of the functions $M_{(ext/int)}(y)$ plays the role of the mass confined
inside the sphere $y<r$.  Constant $C$ can be determined by matching
both sectors of $f(r)$ at
$r=r_0$ that yields  $C=M_{(int)}(r_0)$. This matching is smooth as far as
$\cH(r_0)=-\tilde L(r_0)$, thus the metric (\ref{gKS}) is smooth
everywhere.

By analyzing dependencies
$L(r)$ and $\cH(r)$, Figs. 4 and 5., one sees that there is a core region
($r<0.6 r_0$)
which does not give contribution to mass at all. The space is flat in this
core since $f(r)\approx 0$ there. The region $0.6\le r/r_0 \le 1$ is a
transitional one with a dual magnetic phase, whereas the region $1\le r/r_0
\le 1.6$ is one with an electric phase. In the region $r>1.6 r_0$ we
have $L_F\sim 1$,  it is practically a Coulomb phase.

Let us estimate characteristic scale $r_0$ for the \AG type of nonlinearity
(\ref{AGH}). Radius $r_0$ is determined by the equation $H_P=0$, or by
the root of equation $\zeta \tanh (\zeta)=2$.
It yields $\zeta=2.066$, and
\be
r_0=\frac {2^{1/4} q^2}{2m \zeta}=.48 r_e,
\label{r0ned}
\ee
where  $r_e=q^2/2m$ is a classical electromagnetic radius.

For comparison, in the Born-Infeld theory (in our notations), Lagrangian is
\be
\cL_{BI}=\frac 1{b^2} (\sqrt{1+2b^2 F}-1),
\label{BI}
\ee
and has a dual Lagrangian
\be
\cH_{BI}=\frac 1{b^2} (1- \sqrt{1-2b^2 P}).
\label{DBI}
\ee
The typical NED branches are absent here, and the sphere $H_P=0$ shrinks
to the point $r=0$. Therefore the considered above type of solutions
is absent in Born-Infeld theory.

Black holes in Born-Infeld theory were
actively studied (see \cite{gibras2} which includes the original
references).
The Born-Infeld \bh solution has finite total energy
$E=m\approx1.236 e^2/r_0$ and a smooth distribution of effective charge
$\rho_{eff}=\frac1{4\pi}div \E$ in the region $\sim r_0 =\sqrt {e b}$ .
Parameter $b^{-1}=e/r_0^2$ characterizes a critical value of the
electromagnetic field.

Considering particlelike solutions, one sees that, similarly to NED,
the effective radius of the elementary particle in Born-Infeld theory is
close to classical electromagnetic one
\be
r_0 =2.472 r_e \sim 10^{-13} cm.
\label{r0BI}
\ee
However, in  contrast to NED, the Born-Infeld solution has a
$\delta$-function source at the origin, leading to a conical singularity in
curvature. Therefore the Born-Infeld theory, as well as NED, has classical
radius $r_e$ as one of the characteristic scales. However, regularizing ( or
smearing ) the \bh singularities, both the theories display that their
{\it minimal} ( space ) scale is extended up to Planck sizes.

On the other hand, the Born-Infeld theory can be derived from open string
theory \cite{Ts,gibras2}, and parameter $b$ takes in this case the value
$b=2\pi \alpha ^\prime =T^{-1}$ defined by Regge slope $\alpha ^\prime$.
On the level of particlelike solutions, it does not contradict with
$b^{-1}\sim e/r_0^2$, leading to approximate value of the fine structure
constant $\alpha \approx \frac 1 {44}$ ( see, e.g, \cite {gibras2}).
However, in the modern superstring theory tension $T_{(ss)}$ defines the
critical scale for all field strength, including curvature,  and
characteristic radius $ r_{crit}=\sqrt {\alpha^\prime_{(ss)}}$ corresponds
to the Planck scale.
It shows that the scale range of nonlinearity in NED action is much larger
than the Planck scale that governs higher order corrections in curvature.

\section{Concluding remarks}
\noindent The \AG solution contains distribution of
electric charges $J^\n_{(e)}= \nabla_\m F^\mn$ near the sphere $r=r_0$ and
in
the center. In the presented modified solution the charges
inside the sphere $r=r_0$ are magnetic $J^\n_{(m)}= \nabla_\m \star \tilde
F^\mn$ fulfilling the Dirac idea on dual electrodynamics \cite{Dir}. These
charges are confined inside the sphere  $r=r_0$, forming a magnetic vacuum
polarization. For the external observer this solution exhibits only electric
charge.
Since vacuum of the core expels electric charges one can speculate that
it possesses superconducting properties, whereas the vacuum of the external
observer demonstrates the dual superconductivity, expelling the magnetic
charges.
The considered matching of the electric and magnetic phases corresponds to a
sharp phase transition between these vacua ( London limit of
superconductivity).

The similar baglike structures with a
phase transition to dual superconductivity ( approximate solutions ) were
recently discussed for regular rotating black holes to supergravity
\cite{Bag}. It was assumed that superconductivity is caused by the Higgs (
chiral ) fields forming two dual ( supersymmetric ) vacuum states and a
smooth domain wall interpolating between them.
It is remarkable that nonlinear electrodynamics allows one to
get {\it exact and self-consistent} solutions of
this sort belonging to the Kerr-Schild class of metrics. However, it could
seem wonderful since the Higgs fields providing the vacuum states and phase
transition are absent here.
On the other hand, the NED Lagrangian $L(F)$ can be represented in the form
$(\frac {L(F)}{F})F$ showing that the factor
$\frac {L(F)}{F}$ can be identified as a dilaton ( corresponding to a
metric on the moduli space in supergravity). It allows one to suppose that
NED can be assumed as a limiting case of supergravity when the potential
is very sharp and chiral fields are dropped out. The presented exact
solution
from NED shows that dilaton has to be very essential in the baglike models.

A parallelism between the above baglike models and the field models of
dual superconductivity should also be mentioned.
Indeed, there are known the dual superconductivity models of two type:
the models with Higgs fields, leading to solutions with a soft
phase transition ( finite thickness of transitional layer ),
see, e.g.,  \cite{Ch,MS}, and the similar to NED models without Higgs
fields,
see, e.g., \cite{SW,KIK}, demonstrating the sharp phase transition connected
with the London limit of superconductivity.
The field models of the first type represent in fact a truncation of the
four-dimensional supergravity, while the models of the second type
represent,
apparently, its singular limit. \fn{Note that effective Higgs fields can
also
be introduced in the models of the second type \cite{KIK}.}  However, these
questions demand more detailed consideration elsewhere.

  It would also be very interesting to look on NED from a high dimensional
point of view, getting it  as a result of compactification or as a four
-dimensional slice of a multidimensional spice-time in the spirit of early
brane world work \cite{GW}.

Finally, taking into account nontrivial topology of the Kerr geometry, its
exceptional character with respect to dual rotations \cite{SGP} and
relationships to the spinning particlelike solutions ( see, e.g., \cite{Bur}
and references therein ), the treatment of corresponding regular {\it
rotating}
solutions is very important and could bring some surprising results. In
particular, the appearance of a closed vortex string on the place of the
Kerr
singular ring is expected.

\section*{Acknowledgments}
 We are thankful to I. Dymnikova for useful discussions at the early stage
of
this work and to K. Bronnikov for presenting his paper, very useful
discussions, and comments.

\section*{Appendix. The Kerr-Schild form of metric in the case of spherical
symmetry}
The Kerr-Schild form of metric \cite{DKS,behm}
in the case of spherical symmetry is
\be
g_\mn=\eta_\mn + 2h k_\m k_\n , \label{AgKS}
\ee
where $\eta_\mn$ is the auxiliary Minkowski metric, $h=f(r)/r^2$, and
$k_\m = \{1,{\bf n}\}$.
In the Kerr angular coordinates $\{t, r,\theta, \phi \}$ we have
\begin{equation} g_{\mn} = \left( \begin{array}{cccc} 2h-1&2h&0&0\\
2h&1+2h&0&0 \\
0&0&r^2&0 \\
0&0&0&r^2\sin ^2 \theta
\end{array} \right),
\label{gK}
\end{equation}
where $h(r)=f(r)/r^2$,
$\sqrt {-g} =r^2\sin\theta$, and the contravariant form
of metric is
\begin{equation}
g^{\mn} =
\left( \begin{array}{cccc}
-(1+2h)&2h&0&0 \\
2h&1-2h&0&0 \\
0&0&1/r^2&0 \\
0&0&0&(r^2 \sin ^2 \theta) ^{-1}
\end{array} \right),
\label{gcontK}
\end{equation}
The energy-momentum tensor for the generalized
Kerr-Schild class of metrics is \cite{behm,Bag}
\be
T_\m^\n = (8\pi)^{-1}{\it diag}\{-2G,-2G,D+2G,D+2G\}, \label{ATKS}
\ee
where
\be
D= -f''/r^2, \quad G= (f/r)'/r^2, \label{ADG}
\ee
 the prima means $ d /d r$, and energy density is $\rho=2G/8\pi$.

\end{document}